\documentclass[preprint, showpacs,preprintnumbers,amsmath,amssymb]{revtex4}
\usepackage{graphicx}
\usepackage{dcolumn}
\usepackage{bm}
\usepackage{amsfonts}%
\usepackage{amsmath}

\begin{document}


\title{Geometrical phase driven predissociation: Lifetimes of 2$^2A'$ levels of H$_3$}
\author{Juan Blandon and Viatcheslav Kokoouline}
\affiliation{Department of Physics, University of Central Florida, Orlando, Florida 32816, USA}

\date{\today}

\begin{abstract}

We discuss the role of the geometrical phase in predissociation dynamics of vibrational states  near a conical intersection of two electronic potential surfaces of a $D_{3h}$ molecule. For quantitative description of the predissociation driven by the coupling near a conical intersection, we developed a method for calculating lifetimes and positions of vibrational predissociated states (Feshbach resonances) for X$_3$ molecule. The method takes into account the two coupled three-body potential energy surfaces, which are degenerate at the intersection. As an example, we apply the method to obtain lifetimes and positions of resonances of predissociated vibrational levels of the 2$^2A'$ electronic state of the H$_3$ molecule. The three-body recombination rate coefficient for the H+H+H $\to$ H$_2$+H process is estimated.

\end{abstract}


\maketitle

Non-Born-Oppenheimer interaction near a conical intersection (CI) between two potential energy surfaces (PES) of polyatomic molecules plays an important role in dynamics of the molecules (see, for example, review \cite{yarkony96}). At the exact point of intersection, the interaction becomes infinite and the Born-Oppenheimer adiabatic (BOA) approximation breaks down. Therefore, the description of the nuclear motion of the molecule with energies around or above the energy of a CI should explicitly take into account the two intersecting PESs coupled by the non-adiabatic interaction. 

\begin{figure}
\includegraphics[width=14cm]{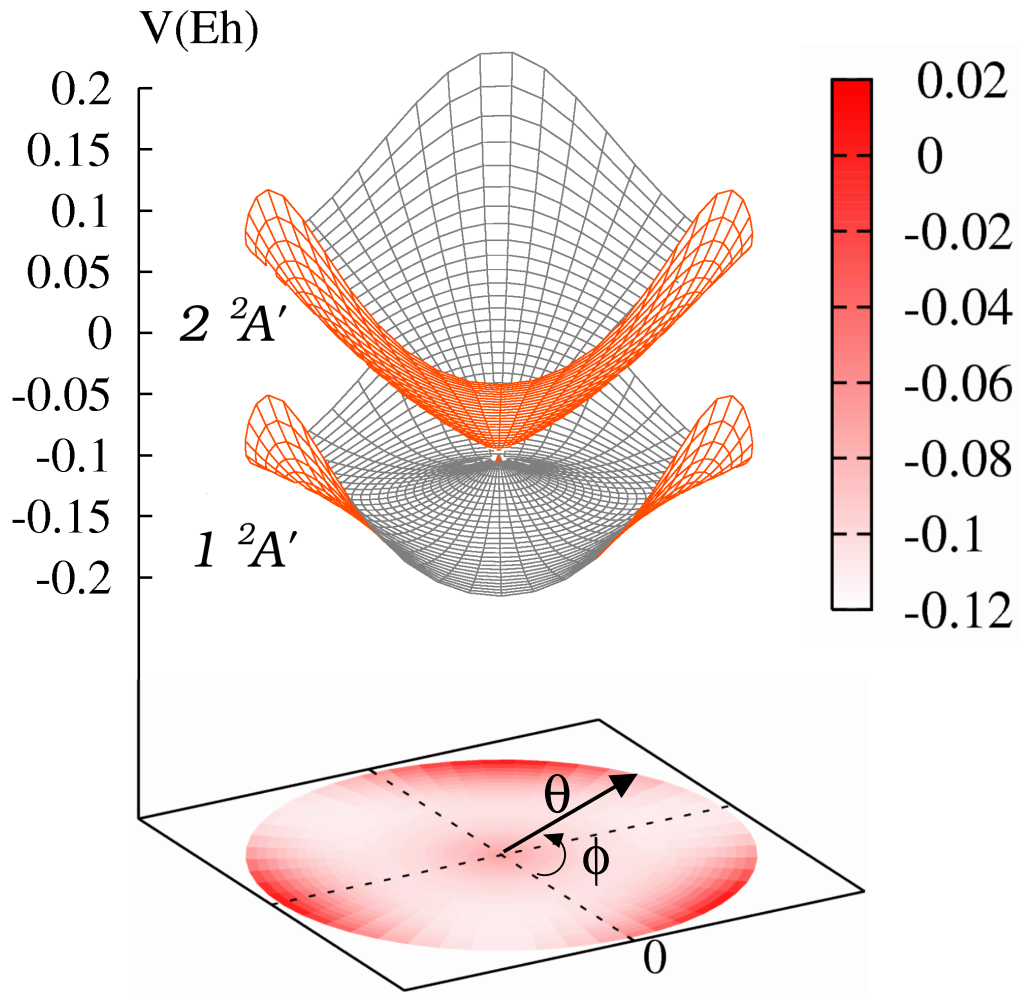}
\caption{(Color online) The two lowest {\it ab-initio} potential energy surfaces of H$_3$ shown as functions of hyperangles $0 \le \theta \le \pi/4$ and $0 \le \phi \le 2\pi$ for a fixed hyper-radius $\rho=2.5\ a_0$.  The projection at bottom of the figure corresponds to 1$^2A'$ PES.}
\label{fig:truhlar_pots}
\end{figure}

One of the simplest processes involving the interaction near a CI is the H+H$_2(v',j')\to$ H+H$_2(v,j)$ scattering process (and its isotopic variants) that has been extensively studied in theory and experiment \cite{kliner90,neuhauser92}. The related processes are the collisional dissociation of H$_2$ and its inverse, the three-body recombination of hydrogen: H+H+H$\to$ H$_2(v,j)$+H, which is responsible, for example, for the formation of the first generation of stars \cite{flower07}. The  processes are governed by the two lowest molecular potential surfaces 1$^2A'$ and 2$^2A'$ of H$_3$ (see Fig. \ref{fig:truhlar_pots}). The lowest 1$^2A'$ potential is repulsive and leads to the H$_2$+H dissociation. The dissociation limit of the  2$^2A'$ potential correlates with the H+H+H breakup and has a number of quasi-bound vibrational levels, which are predissociated towards to the H$_2$+H dissociation due to the coupling near the CI. 
The predissociation of 2$^2A'$ vibrational states has been studied by Kupperman and collaborators \cite{lepetit07} using a time-independent scattering framework, the time-delay analysis, and a combination of Jacobi and hyperspherical coordinates. The calculation has been done with the two-channel diabatic potential of H$_3$ with non-diagonal diabatic couplings obtained directly from first derivatives of \textit{ab initio} BOA electronic wave functions of the two interacting states. In another study by Mahapatra and K\"oppel \cite{mahapatra98b}, the time-dependent approach was employed by propagating a wave packet that is initially placed on the 2$^2A'$ PES; the lifetimes are then derived from the autocorrelation function. In Ref. \cite{mahapatra98b}, the authors also used a diabatic representation of the coupled H$_3$ potential, but the diabatization is done differently than in Ref. \cite{lepetit07} and does not require explicit calculation of first derivatives of the BOA electronic wave functions. The lifetimes obtained in the two studies are significantly different.

In this study, we suggest a general theoretical method to describe the nuclear dynamics involving two molecular potentials coupled by a non-adiabatic interaction near the CI. As an example, we calculate lifetimes of several 2$^2A'$ predissociated vibrational levels of H$_3$. The method can be used for other small polyatomic molecules where the CI plays an important role. 
There are two main ingredients in the proposed method: (1) The diabatization procedure is made in a way that accounts for the Jahn-Teller coupling between the 1$^2A'$ and 2$^2A'$ molecular states and the $D_{3h}$ symmetry of the total vibronic wave function. (2) Nuclear dynamics is described by Smith-Whitten hyperspherical coordinates \cite{johnson83}, adiabatic separations of hyperangles and hyper-radius along with the slow-variable discretization (SVD) \cite{kokoouline06,blandon07}, and a complex absorbing potential (CAP) to obtain resonance lifetimes.

\begin{figure}
\includegraphics[width=15cm]{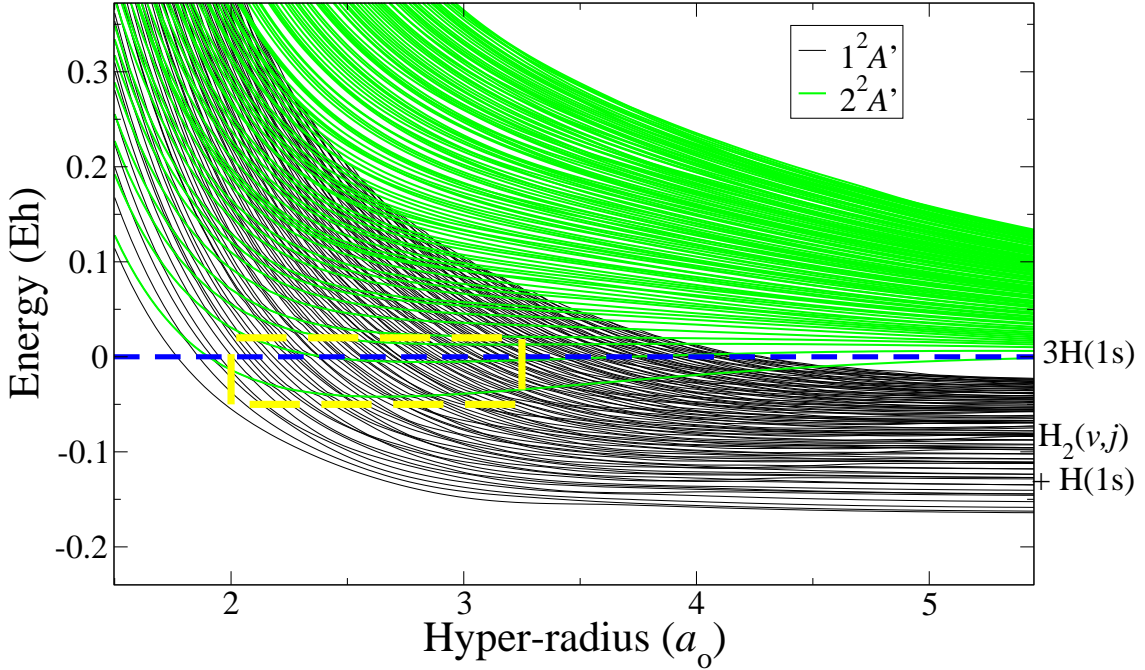}
\caption{(Color online) Hyperspherical adiabatic potential curves, obtained from uncoupled $1^2A'$ (black curves) and  $2^2A'$ (green curves) PESs of H$_3$. Different dissociation limits for the $1^2A'$ family correspond to different $v$ and $j$. Here, we only show the curves of the $A_1$ irreducible representation. }
\label{fig:H3_adiabats1}
\end{figure}

\begin{figure}
\includegraphics[width=15cm]{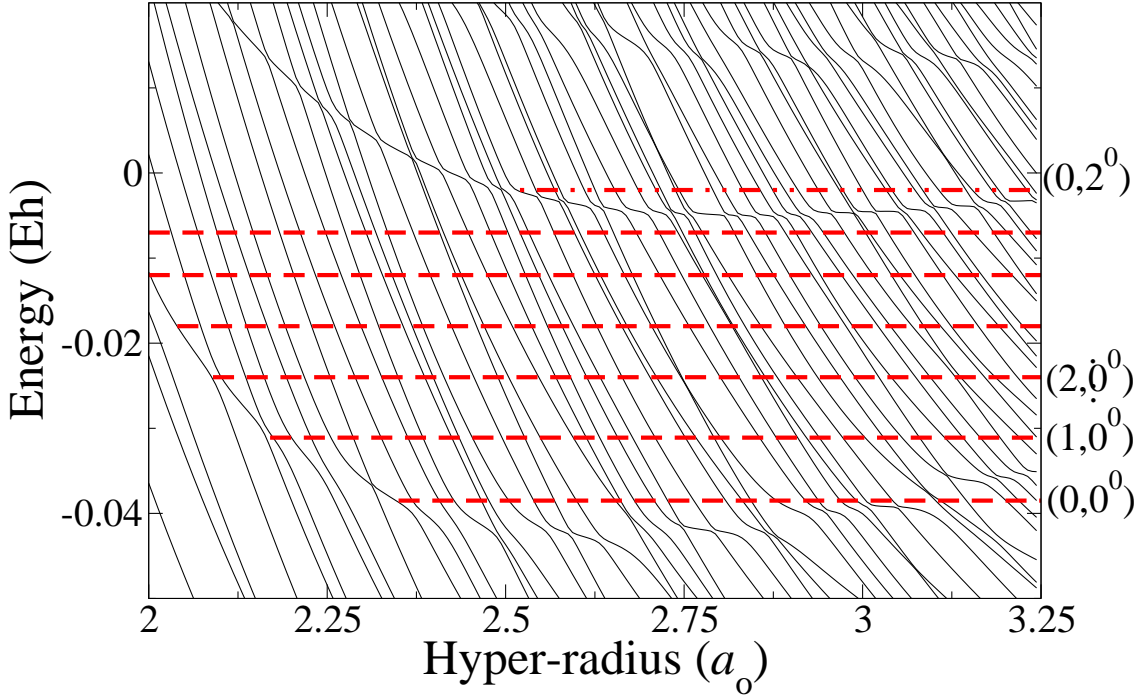}
\caption{Close up look at avoided crossings in HSA curves when the PESs are represented by Eq. (\ref{eq:couple_pots}). The figure corresponds to the frame shown in Fig. \ref{fig:H3_adiabats1}, where HSA curves are calculated from the uncoupled {\it ab initio} PESs. The horizontal dashed lines show the positions of predissociated $2^2A'$ levels.}
\label{fig:avoided_crossings}
\end{figure}

\begin{figure}
 \includegraphics[width=15cm]{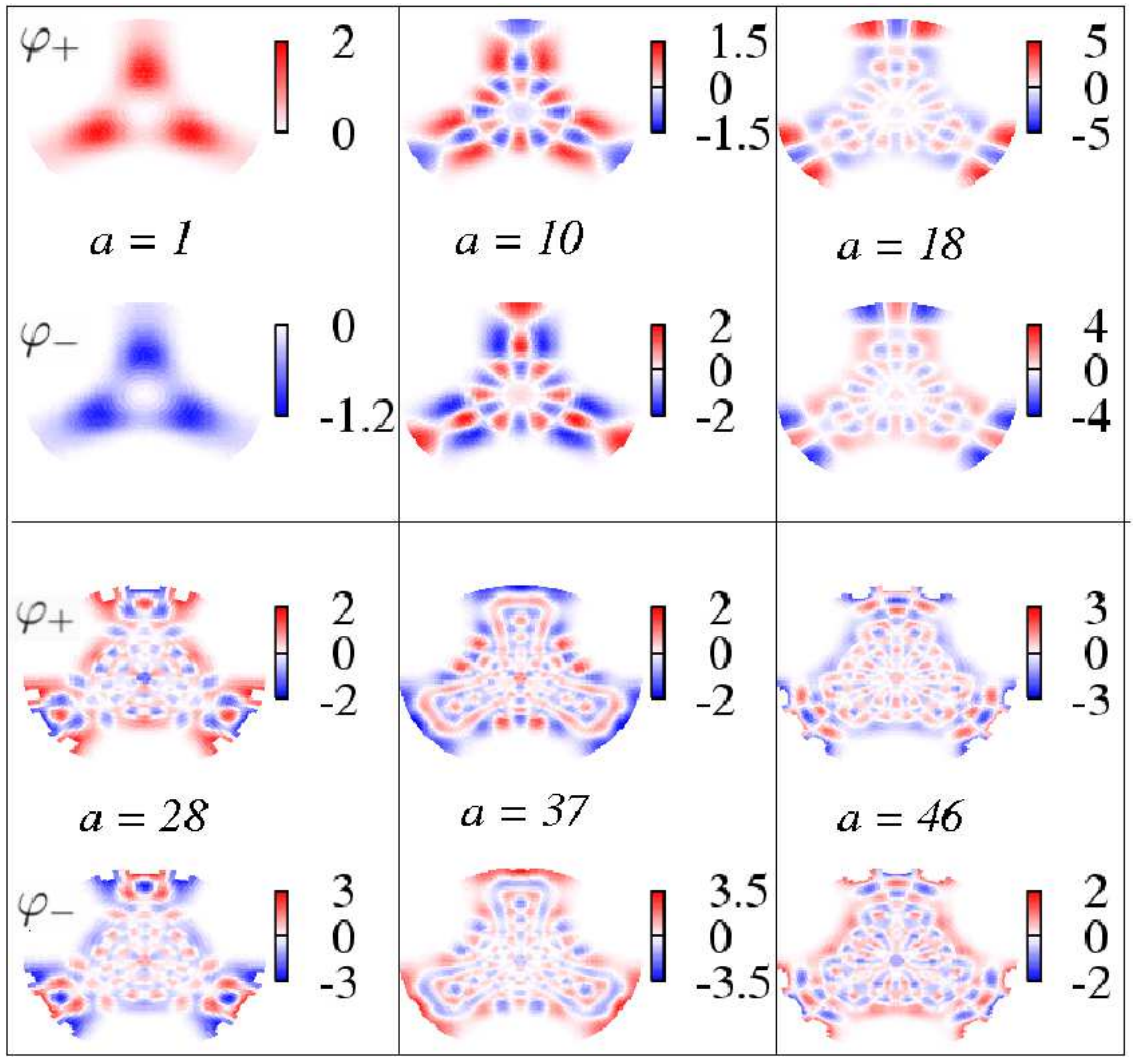}
\caption{\label{fig:hpa_wfs} (Color online) Wave functions of HSA states $\varphi_{a,j}$ of the $A_1$ irreducible representation as functions of the hyperangles $\theta$ and $\phi$ for $\rho_j=2.5\ a_0$ and several different $a$. Each wave function has two components, $\varphi_+$ and $\varphi_-$ corresponding to the two channels of the potential $\hat V$. The relationship between the hyperangles and the three-body configurations they represent is mapped on Fig. 6 of Ref. \cite{kokoouline03b}. }
\end{figure}

{\bf Treatment of nuclear dynamics.} To treat the vibrational dynamics in hyperspherical coordinates $\rho,\theta,\phi$, we represent the vibrational wave function $\psi(\rho,\theta,\phi)$ as the expansion $\psi(\rho,\theta,\phi)=\sum_{k}y_k(\rho,\theta,\phi)c_k$ in the basis of non-orthogonal basis functions $y_k=\pi_j(\rho)\varphi_{a,j}(\theta,\phi)$. In this expression, $\varphi_{a,j}(\theta,\phi)$ is the hyperspherical adiabatic (HSA) state calculated at a fixed hyper-radius $\rho_j$,
\begin{equation}
\label{eq:Had}
H_{\rho_j} \varphi_{a,j}(\theta,\phi)=U_a(\rho_j)\varphi_{a,j}(\theta,\phi)\,,
\end{equation}
with the corresponding eigenvalue $U_a(\rho_j)$.  $H_{\rho_i} $ is the two-channel vibrational Hamiltonian with the hyper-radius fixed at $\rho_j$. If $\rho_j$ changes continuously, energies $U_a(\rho_j)$ and the wave functions $\varphi_{a,j}(\theta,\phi)$ form the HSA curves $U_a(\rho)$ and channel functions $\varphi_{a,\rho}(\theta,\phi)$, respectively, (see Figs. \ref{fig:H3_adiabats1}, \ref{fig:avoided_crossings}, \ref{fig:hpa_wfs}). In the basis $y_k$, the Schr\"odinger equation for the total wave function $\psi(\rho,\theta,\phi)$ takes the form of a generalized eigenvalue problem for coefficients $c_k\equiv c_{ja}$
\begin{eqnarray}
 \label{eq:gen_eigen}
 \sum_{j',a'}\Big[\langle\pi_{j} |-\frac{\hbar^2}{2\mu}\frac{d^2}{d\rho^2}|\pi_{j'}\rangle{\cal O}_{ja,j'a'}+\langle\pi_{j}|\hat U_a(\rho)|\pi_{j'}\rangle\delta_{aa'}\Big]c_{j'a'} \nonumber\\ 
=E_n ^{vib} \sum_{j',a'}\langle\pi_{j}|\pi_{j'}\rangle{\cal O}_{ja,j'a'}c_{j'a'}\,.
\end{eqnarray}
with overlap matrix elements
\begin{equation}
\label{eq:couplings_svd}
 {\cal O}_{ja,j'a'}=\langle\varphi_{a,j}(\theta,\phi)|\varphi_{a'j'}(\theta,\phi)\rangle ,
\end{equation}
that replace the familiar non-adiabatic couplings between the $\varphi_{a,\rho}(\theta,\phi)$ channels. This way of representing the non-adiabatic hyperspherical couplings provides an important advantage over the familiar method of dealing with the couplings using the first and second derivatives of $\varphi_{a,\rho}(\theta,\phi)$ with respect to hyper-radius. In order to obtain lifetimes of predissociated vibrational levels, we place a CAP at a large value of the hyper-radius (for details, see Ref. \cite{blandon07}). The total three-body rotational angular $J$ momentum is 0 in the present calculation.

{\bf Diabatic basis for the coupled H$_3$ potential.} We use the adiabatic {\it ab initio}  1$^2A'$ and 2$^2A'$ PES from Ref. \cite{varandas87}, which will be referred as $V_1$ and $V_2$. The dependence of the potentials on the two hyperangles is shown in Fig. \ref{fig:truhlar_pots}. In our approach, we also use a diabatic representation of the interaction potential. However, we derive the non-diagonal diabatic coupling elements from the {\it ab initio} PESs without calculating them explicitly as derivatives of Born-Oppenheimer electronic states. This way of representing the non-adiabatic couplings has been very successful in diatomic molecules: If the two {\it ab initio} PESs have an isolated avoided crossing, the vibrational dynamics is well represented by a $2\times 2$ diabatic potential with the geometry-independent non-diagonal matrix  element equal to the half of the splitting between the  {\it ab initio} PESs at the avoided crossing. The diabatic non-diagonal coupling in our model is derived in the following way. 

The two 1$^2A'$ and 2$^2A'$ electronic states become degenerate at the equilateral configuration and should be referred to as two components of the $E'$ irreducible representation of the $D_{3h}$ symmetry group. It is convenient to use the basis functions $\vert E_+\rangle$ and $\vert E_-\rangle$ in the two-dimensional $E'$ space \cite{douguet08,longuet61}.  Although the electronic states  1$^2A'$ and 2$^2A'$ for clumped nuclei are classified according to the $C_s$ symmetry group, the vibronic states of H$_3$  should be classified according to the $D_{3h}$ group. The most general form of the diabatic potential in the basis of $\vert E_\pm\rangle$ for an arbitrary geometry is 
\begin{eqnarray}
\label{eq:couple_pots}
\hat V = \left ( \begin{array}{cc}
A & C e^{if} \\ 
C e^{-if} & A \end{array}\right ),
\end{eqnarray}
where $A$, $C$, and $f$ are real-valued functions of the three hyperspherical coordinates \cite{kokoouline03b}. The diagonal elements $A$ are the same because of the degeneracy of $E_\pm$ states. The functions $A$ and $C$ transform in the $D_{3h}$ symmetry group  according to the $A_1$ representation, and $f$ has the following property under the $C_3$ symmetry operator: $C_3 f=f+2\pi/3$. The $A$ and $C$ functions are uniquely determined from the 1$^2A'$ and 2$^2A'$ PESs: $A=(V_1+V_2)/2$ and $C=(V_2-V_1)/2$. The actual form of function $f$ can be derived near the CI: It is equal to the phase of the asymmetric normal mode distortion \cite{longuet61,kokoouline03b}. Although this form of $f$ is derived near the CI, we will use it everywhere. This is justified because (1) the transition between adiabatic states occur only near the CI, (2) far from the CI the phase factor $e^{\pm if}$ does not play a role as long as the symmetry property mentionned above is satisfied.

The vibrational wave functions $\psi$ obtained from Eq. (\ref{eq:gen_eigen}) have two components $\psi_\pm$ corresponding to the two diabatic $E'_\pm$ basis functions. In the adiabatic basis, corresponding to the 1$^2A'$ and 2$^2A'$ electronic states, the two components $\psi_{1,2}$ of the $\psi$ function have the form
\begin{eqnarray}
\psi_{1} = (\psi_- e^{if/2} + \psi_+e ^{-if/2})/\sqrt{2}\,,\nonumber\\
\psi_{2} = i(\psi_- e^{if/2} - \psi_+e ^{-if/2})/\sqrt{2}\,.
\end{eqnarray}
After applying the $C_3$ symmetry operator three times, the molecule returns back to its original position, however the components $\psi_{1,2}$ change sign  $\psi_{1,2}\to-\psi_{1,2}$ because $f\to f+2\pi$. It is a well-known property of adiabatic states in the presence of CI, which is often referred as geometrical or Berry phase effect. Because the adiabatic electronic wave functions $\vert 1\rangle$ and $\vert 2\rangle$ also change sign, the total vibronic wave function 
\begin{eqnarray}
\Psi = \psi_{1} \vert 1\rangle +\psi_{2} \vert 2\rangle =\psi_{+} \vert E_+\rangle +\psi_- \vert E_-\rangle
\end{eqnarray}
is unchanged after the identity operator $C_3^3$ is applied.

{\bf Results.}  After solving the hyperangular part of the three-body Hamiltonian, Eq. (\ref{eq:Had}), for each BOA PES separately, we obtain two uncoupled 'families' of adiabatic potentials $U_a(\rho)$, which are shown in Fig. \ref{fig:H3_adiabats1}. The curves belonging to the different families can cross. When the coupling is turned on between the two electronic states, the crossings turn into avoided crossings, which is demonstrated in Fig. \ref{fig:avoided_crossings}. The figure shows the  HSA curves obtained by solving Eq. (\ref{eq:Had}) with the diabatic electronic potential of Eq. (\ref{eq:couple_pots}). Far from the (avoided) crossings the calculations with coupled and uncoupled electronic states produce almost the same HSA states. The coupling changes the HSA states at (avoided) crossings only. It is also worth to mention, that if the phase factors $e^{\pm if}$ in Eq. (\ref{eq:couple_pots}) are neglected, the dynamics described by such diabatic potential is {\it exactly} the same as the dynamics with the uncoupled BOA PESs. It is because the transformation diagonalizing the operator $\hat V$ is independent of the nuclear coordinates if $f\equiv0$.  Fig. \ref{fig:hpa_wfs} shows the vibrational HSA functions $\varphi_{a,j}$ for several adiabatic states calculated with the coupled potential $\hat V$ of Eq. (\ref{eq:couple_pots}). 

As Figs. \ref{fig:H3_adiabats1} and \ref{fig:avoided_crossings} show, the ground and first excited HSA potentials of the $2^2A'$ family have minima and can have vibrational levels that are pre-dissociated due to the coupling with the $1^2A'$ state. Although each vibrational level has components from all the HSA curves shown in Fig. \ref{fig:H3_adiabats1} (or Fig. \ref{fig:avoided_crossings}), only one component is dominant. Thus, the vibrational levels can be characterized by (1) the dominant component $a$ and by (2) the number $v_1$ of quanta along the hyper-radius. In addition, each HSA curve in the  $2^2A'$ family can be characterized by the number $v_2$ of quanta in the hyper-angular space and the number $l_2$ of those $v_2$ quanta along the cyclic hyperangular coordinate $\phi$. Therefore, each predissociated level can be numbered with the triad $\{v_1,v_2^{l_2}\}$ similar to the normal mode notations for $C_{3v}$ molecules. 

\begin{table}[tbp]
\begin{tabular}{|p{1.3 cm}|p{2.8 cm}|p{2.5 cm}|p{2.5 cm}|}
\hline
  $\{v_1, v_2 ^{l_2}\}$ &$E_r, \tau$; this work  & $E_r, \tau$; Ref. \cite{lepetit07}& $E_r, \tau$; Ref. \cite{mahapatra98b}  \\
\hline
 $\{0, 0 ^{0}\}$  & $ -3.85$, $ 13. $ & n.a.& $-3.74$, $\sim 3 $ \\
\hline
 $\{1, 0 ^{0}\}$  &    $-3.11$, $ 13. $ & n.a.&  $-3.01$, $\sim 3 $\\
\hline
 $\{2, 0 ^{0}\}$  &    $-2.4$, $ 14. $  & n.a.& $-2.32$, n.a.\\
\hline
  $\{3, 0 ^{0}\}$ &    $-1.8$, $ 14. $ & n.a.& $-1.70$, n.a. \\
\hline
  $\{4, 0 ^{0}\}$&    $-1.2$, $ 16. $ & $-1.19$, $\sim 15. $ &  $-1.14$, n.a.\\
\hline
  $\{5, 0 ^{0}\}$ &    $-0.7$, $ 18.$  & $-0.42$, $\sim 17. $& $-0.65$, n.a.\\
\hline
  $\{0, 2 ^{0}\}$  &    $-0.2$, $ 130. $  & n.a.& $-0.22$, $\sim 4.5 $\\
\hline
\end{tabular}
\caption{Positions, $E_r$ (in units $10^{-2}$ Eh) and lifetimes, $\tau$ (in fs) of pre-dissociated $2^2A'$ vibrational levels. Energies are relative to the H($1s$) + H($1s$) + H($1s$) dissociation. }
\label{table:res_energies}
\end{table}

Below the H+H+H dissociation limit there are only two $\{v_10^0\}$ and $\{v_12^0\}$ series of  $A_1$ predissociated levels. Their positions and lifetimes are given in Table \ref{table:res_energies}. Lifetimes for the two series are very different. The reason for the difference is that the avoided crossings are significantly wider for the $0^0$ curve than for the $2^0$ curve (see Fig. \ref{fig:avoided_crossings}). The table also compares the obtained results with two other studies \cite{lepetit07,mahapatra98b} of the $2^2A'$ predissociated levels. In Ref. \cite{lepetit07} the non-diagonal couplings in the diabatic basis are obtained from the accurate {\it ab initio} non-Born-Oppenheimer couplings between the $1^2A'$ and $2^2A'$ states, but not from PESs as in this study. The agreement with our study for the lifetime and energy of the $\{40^0\}$  resonance is very good as well as for the lifetime of the $\{50^0\}$ level. The agreement for the energy of the $\{50^0\}$ level is not as good probably due to the somewhat special character of the $\{50^0\}$ level: its hyper-radial wave function extends to relatively large values $\sim 7 a_0$ of hyper-radius (and internuclear distances $\sim 5 a_0$). In Ref. \cite{mahapatra98b}, the diabatization procedure was based on PESs similarly as it is made in the present study, but with one important difference: the diabatic electronic states in \cite{mahapatra98b} are non-equivalent. It means that the $D_{3h}$ character of the vibronic wave functions in Ref. \cite{mahapatra98b} is broken. The lifetimes obtained in Ref. \cite{mahapatra98b} are significantly different from the present values and values of Ref. \cite{lepetit07}. The disagreement is attributed to the choice of the diabatization procedure in \cite{mahapatra98b} that does not respect the degeneracy of the diabatic electronic wave functions.

It is worthwhile to note the relatively small number $N_\rho=96$ of hyper-radial points $\rho_j$ required to obtain converged lifetimes and positions of resonances considering the large number of sharp avoided crossings (see Fig. \ref{fig:avoided_crossings}). It is an advantage of using the SVD procedure rather than the familiar $\langle\varphi_a\vert\partial^2/\partial \rho^2\vert\varphi_{a'}\rangle$ and $\langle\varphi_a\vert\partial/\partial \rho\vert\varphi_{a'}\rangle$ non-adiabatic couplings between the HSA states: It is not necessary to calculate the derivatives on a fine grid of hyper-radius near avoided crossings to describe the couplings locally. Instead, in the SVD approach, the overlap matrix elements $ {\cal O}_{ia,i'a'}$  between the adiabatic states in Eq. (\ref{eq:couplings_svd}) account globally for the non-adiabatic hyperspherical couplings. It allows us to reduce significantly the computational task.

The obtained lifetimes and positions of the resonances can be used to estimate the three-body recombination rate coefficient $k_3$ for the H+H+H$\to$H$_2$+H reaction. Using the probability $P \sim 1/(\tau\omega)$ of the transition from $2^2A'$ to $ 1^2A'$, where $\omega$ is the frequency of oscillation in the $2^2A'$ potential, and using the formula for $k_3$ from Ref. \cite{esry99}, taking sum over total angular momentum $J$ up to such $J_{max}$ that the lowest $2^2A'$ vibrational level moves above the H+H+H dissociation, and also taking the sum over all three types of vibrational levels ($A_1$, $A_2$, and $E$), we obtain value $k_3\sim 2\times 10^{-30}$cm$^6/$s at 300~K with estimated error of about 50\%. This value is in good agreement with other estimation \cite{flower07}, $k_3\sim 2.2\times 10^{-30}$cm$^6/$s at 300~K, derived using the detailed balance principle from the rate of the H$_2$+H$\to$H+H+H processes. Rigorous calculations for $k_3$ combining the developed model for the H$_3$ potential, the SVD approach, and the three-body R-matrix approach are under way.

Concluding, we would like to stress that we (1) suggested a model diabatic two-channel molecular potential that represents correctly the interaction between vibronic states near a CI. The non-diagonal as well as diagonal matrix elements of the potential are extracted from the {\it ab initio} adiabatic PESs of the molecule. There is no need to use {\it ab initio} non-adiabatic couplings between PESs to construct the diabatic potential. (2) We applied the model potential for calculation of lifetimes of $2^2A'$ pre-dissociated levels in H$_3$. (3) Using the obtained resonances, we derived the rate coefficient for the three-body recombination of three hydrogen atoms. The developed techniques combining the model diabatic potential and the numerical method can also be applied to other systems where a CI is expected to play a role.

We would like to thank the Donors of the American Chemical Society Petroleum Research Fund, the National Science Foundation under Grant No. PHY-0427460 for an allocation of NCSA and NERSC supercomputing resources (project \# PHY-040022), and the Florida Education Fund McKnight Doctoral Fellowship for support of this research.


\end{document}